\documentclass[journal,draftcls,onecolumn,12pt,twoside]{IEEEtranTCOM}

\normalsize
\usepackage{url}
\usepackage{tabularx,ragged2e,booktabs}
\usepackage{slashbox}
\usepackage{subcaption}
\usepackage{ifpdf}
\usepackage{cite}
\usepackage{algorithm}
\usepackage{amsmath, amsthm, amssymb, amsfonts}
\usepackage{graphicx}
\usepackage{algpseudocode}
\usepackage{epstopdf}
\usepackage{amsmath}
\usepackage{array}
\usepackage[usenames,dvipsnames]{color}

\DeclareMathOperator*{\argmax}{arg\,max}

\usepackage{fixltx2e}

\begin{document}
%
\title{Model-Based Detector for SSDs in the Presence of Inter-cell Interference}

\author{\IEEEauthorblockN{Hachem Yassine, Mihai-Alin Badiu, and Justin P. Coon, \IEEEmembership{Senior Member, IEEE}}

\thanks{The authors are with Department of Engineering Science, University of Oxford, United Kingdom, OX1 3PJ (e-mail: hachem.yassine@eng.ox.ac.uk, Mihai.badiu@eng.ox.ac.uk justin.coon@eng.ox.ac.uk). M.-A. Badiu is also with the Department of Electronic Systems, Aalborg University, Fredrik Bajers Vej 7, Aalborg, 9220, Denmark.}%


}


%


\maketitle

\begin{abstract}
In this paper, we consider the problem of reducing the bit error rate of flash-based solid state drives (SSDs) when cells are subject to inter-cell interference (ICI). By observing that the outputs of adjacent victim cells can be correlated due to common aggressors, we propose a novel channel model to accurately represent the true flash channel. This model, equivalent to a finite-state Markov channel model, allows the use of the sum-product algorithm to calculate more accurate posterior distributions of individual cell inputs given the joint outputs of victim cells. These posteriors can be easily mapped to the log-likelihood ratios that are passed as inputs to the soft LDPC decoder. When the output is available with high precision, our simulation showed that a significant reduction in the bit-error rate can be obtained, reaching $99.99\%$ reduction compared to current methods, when the diagonal coupling is very strong. In the realistic case of low-precision output, our scheme provides less impressive improvements due to information loss in the process of quantization. To improve the performance of the new detector in the quantized case, we propose a new iterative scheme that alternates multiple times between the detector and the decoder. Our simulations showed that the iterative scheme can significantly improve the bit error rate even in the quantized case.

\end{abstract}

%
\IEEEpeerreviewmaketitle

\section{Introduction}
The modern paradigms of machine learning and artificial intelligence were made possible only through the abundance of computational power and  available data. It is estimated that by 2020, $12000$ terabytes of data will be generated per second \cite{DOMO}, while currently, the largest commercial storage device has a capacity of 100 terabytes \cite{Nimbus}. 
Therefore, it is becoming increasingly important to figure out how to store and manage it.
The ideal non-volatile storage system is fast, reliable, power efficient, cheap and dense in capacity. While not perfect, magnetic hard disk drives (HDDs) are cheap and offer a high bit density, and have been popular for mass storage applications, e.g., personal computers and data centers. However, their major limitation is the existence of mechanical parts that can easily break, and limit their access speed \cite{aritome2015nand}. In the last decade, solid state drives (SSDs), which are based on the semiconductor NAND flash memory \cite{masuoka1987new}, has been a strong competitor that is constantly gaining ground. Being purely electric, NAND flash provides, compared to HDDs, more reliability, lower power consumption, scalability due to Moore's law, and faster access. Over the last 30 years, flash memory evolved to become a standard technology found in a wide variety of applications ranging from consumer electronics to primary memory, e.g., music players, digital cameras, smartphones, personal computers. An attestation to the success of flash memory is that in 2003, the cost of 128 megabytes of flash memory was USD 33, while in 2017, the cost of  128 gigabytes was USD 28, i.e., the megabyte got 1200 times cheaper \cite{FlashPrice}. 

The aggressive cost-per-bit reduction was made possible thanks to several innovations in the industry including size shrinkage and multi-level cells. As a consequence, inter-cell interference (ICI) became a major reliability issue in modern flash memories \cite{grupp2012bleak}. 
In order to maintain virtually error-free storage, it is important to have system level approaches that reduce the probability of errors. One of the most prominent approaches is the use of error-correcting codes (ECC). With the increase in the number of errors, there is a shift from simple, hard-decoded ECC schemes such as BCH and RS codes to the more advanced schemes such as LDPC codes with soft-decoding. The performance of soft-decoding is known to depend on the precision of the read output of the cells and on having an accurate model of the flash channel. The model consists of a probabilistic characterization of the output threshold voltage ($\text{V}_{th}$) of the cells given their inputs. 

In the flash channel modeling literature, the focus is always on the characterization of a single cell. The underlying assumption is that cells are identical and that their outputs are identically and independently distributed according to the conditional probability distribution $p_{Y|X}(y|x)$ where $Y$ is the output of a cell and $X$ is its input. Several works were focused on fitting empirical measurements from  devices to an assumed parametric model of $p_{Y|X}(y|x)$ by minimizing some metric. For example, \cite{lee2013estimation,cai2013threshold} considered a Gaussian distribution, \cite{parnell2014modelling} considered a Normal-Laplace mixture, \cite{luo2016enabling} considered a Student’s t-distribution, and \cite{taranalli2016channel} considered a discrete Beta-binomial distribution. What these works have in common is that they treat interference as an independent noise, that together with the actual noise form an overall independent noise term which the assumed distribution accounts for. Alternatively, when an ICI-aware model is pursued, it is often the case that ICI is first processed out of the outputs via equalization and then the ICI-removed output is characterized \cite{dong2010using,lee2013least,lee2013soft,asadi2014optimal,yassine2016towards}. However, this approach of dealing with ICI is not always feasible as it requires access to the interfering cells in order to equalize their effect on the victim cells.  In solid state drives (SSDs), a file is typically spread across multiple dies/chips that can be read in parallel to increase data throughput. In this case, it is common that the data in the interfering cells are not necessarily requested, hence reading them will incur significant latency and power consumption. In short, ICI equalization is only efficient when a large sequential portion of a flash block is requested by the user.

When equalization is not feasible, ICI must be treated as noise. However, there are two main aspects of ICI that the work treating ICI as noise ignore. The first is that ICI depends on the inputs of the interfering cells and can therefore be more complicated than what can be accounted for by a simple density, e.g., Gaussian. In fact, the empirical investigation of ICI in \cite{cai2013program} has shown that a mixture density better describes its distribution. The second aspect is the fact that by definition, ICI creates a correlation between different cells. The obvious cause of this correlation is that one cell interferes with another, but more importantly, even if two cells do not directly interfere with each other, they can be the subject of interference by a third cell, making them correlated by transition. 

In this paper, we consider both the aforementioned aspects. In particular, we model the flash channel using a hidden Markov model where the hidden states of the channel are defined by the inputs of the aggressor cells. This model differentiates between the independent noise and ICI, and accounts for the correlation between victim cells due to common aggressors. Based on this model, we use the sum-product algorithm to soft-detect the inputs. When the outputs of the detector are used as inputs to the soft LDPC decoder, we show that the adverse effect of ICI can be significantly reduced without accessing the aggressors when the cells' outputs have high precision. The improvement becomes more significant as diagonal coupling increases. However, we show that the proposed scheme is sensitive to output quantization. Hence, to make our scheme realistic, we consider the effect of low-precision quantization of the output. We show that quantization can significantly limit the benefit of our scheme. To mitigate this effect, we propose an iterative scheme that alternates between detection and decoding. We show that the iterative scheme can significantly improve the performance of the LDPC soft decoder even when the quantization precision is low.

The rest of this paper is organized as follows.In section \ref{Basics}, we introduce the basics of flash memory and our model. In section \ref{SingleCellDist}, we derive the output distribution of a single cell. In section \ref{GraphMod}, we present our model-based detector. In section \ref{SImRes}, we present and discuss the simulation results. In section \ref{EffectQuantization}, we discuss the effect of quantization on the performance of the proposed detector and propose an iterative scheme to improve it. The paper concludes in section \ref{Conclusion}.


\begin{figure}[t]
\centering
\includegraphics[scale=0.5]{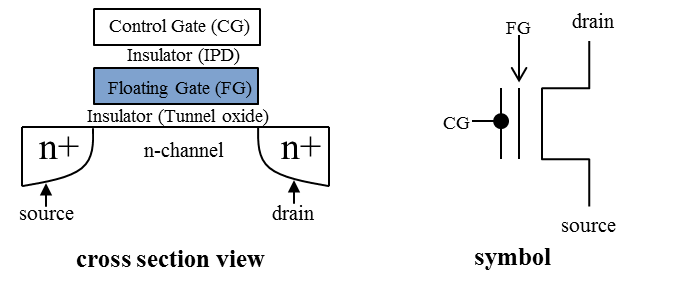}
\caption{Schematic of a floating gate transistor.}
\label{Fig:FGMOS}
\end{figure}
\section{Flash Memory Basics and Model}
\label{flashBasics}
\begin{figure}[t]
\centering
\includegraphics[scale=0.8]{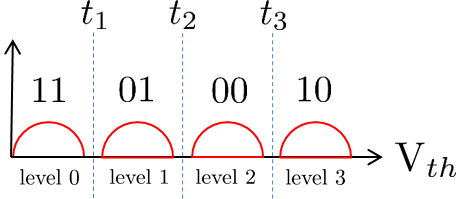}
\caption{Gray mapping of the levels to data bits for MLC.}
\label{Fig:MLCMAPPING}
\end{figure}

\label{Basics}
A planar NAND flash memory block is a two-dimensional grid of floating-gate transistors (cells). The rows are called word-lines (WL), and the columns are called bit-lines (BL).
When a charge is injected into the floating gate, it gives rise to a threshold voltage ($V_{th}$). 
This continuous $V_{th}$ is then quantized into $q$ levels (windows), namely $\mathbb{Z}_q=\{0,...,{q-1}\}$, so that $\log_2 q$ bits of information can be modulated in each cell. Fig. \ref{Fig:MLCMAPPING} shows this quantization with a Gray labelling of the levels for $q=4$. 
Before being programmed, a cell must be erased. The erase operation shifts the cell's $V_{th}$ back to level $0$, while the program operation shifts it up into the level corresponding to the intended symbol (bit sequence). 
When reading it,  the $V_{th}$ of a cell is not readily available, but only a limited precision sensing is possible, i.e., the output is always quantized and a higher precision $V_{th}$ can be obtained with more readings at the expense of increased latency. In the next four sections, we assume that the output is not quantized. In section \ref{EffectQuantization}, we consider the effect of quantization on the performance and then propose an iterative scheme to improve the performance under quantization.
Ideally, the erase operation would push the $\text{V}_{th}$ of all cells to a nominal voltage $v_0$, while the program operation shifts the cell's $\text{V}_{th}$ into one of $q-1$ nominal voltages $v_1,...,v_{q-1}$ representing the $q-1$ program levels. However, the acutal $\text{V}_{th}$ fluctuates around this mean due to circuitry noise, degradation and charge leakage (see Fig. \ref{Fig:MLCMAPPING}) . For simplicity, we abstract the different sources of noise and consider a single noise term that follows a zero-mean, input-dependent Gaussian distribution.

The parasitic capacitance coupling between cells gives rise to ICI, where the voltage shift in one cell induces a proportional shift in its neighbours. A cell is subject to interference only by cells programmed subsequently to it, hence the sequence of programming cells implies which cells interfere with which. In the even-odd bit-line structure (EOBL), even, then odd BLs along each WL are alternately  programmed,  subjecting the odd cells to interference by the even ones. Alternatively, in the all bit-line structure (ABL) all BLs are programmed at the same time, causing no interference to each other. When more than a single bit/cell is stored, the bits belong to different pages. The different pages of a WL can be programmed at the same time (full-sequence), resulting in interference only from the subsequently programmed WL. Alternatively, pages from adjacent (above and below) WLs can be programmed in the intermediary (shadow-sequence), resulting in interference from both sides.
In the rest of this paper, we consider the ABL structure with full-sequence programming of WLs in increasing order. We also assume that the coupling between non-immediate neighbours is negligible. Consequently, each victim cell is subject to ICI from the three neighbour cells on the next WL as shown in Fig. \ref{AnaStruct}.

We denote by $X\in\mathbb{Z}_q$ and $Y\in\mathbb{R}$ the input level and output voltage of a cell, respectively. Hence, it is possible to write the $\text{V}_{th}$ as a communication channel as follows

\begin{equation}
\label{channelmodel}
Y=v_X+Z+W.
\end{equation}
$Z$ is the input-dependent noise term having $p_{Z|X}(z|x)=\mathcal{N}(z|0,\sigma^2_k)$ as its pdf where $\sigma^2_k$ is the noise variance when the input is the $k$th level. $W$ is the ICI that the cell is subject to. Assuming that $m$ cells interfere with the victim cell, $W$ can be written as follows

\begin{align}
\label{eq:ICIexp}
W=\sum_{j=1}^{m}\gamma_{j}\left(Y_{j}^{a}-Y_j^{a,e}\right),
\end{align}
where the superscript `$a$' stands for `aggressor', and the superscript `$e$' stands for `erased'.  $\gamma_j$ is the parasitic capacitance coupling ratio between the $j$th direct aggressor and the victim, $Y_j^{a}$ and $Y_j^{a,e}$ are $\text{V}_{th}$ of the $j$th direct aggressor after being programmed and when it is erased, respectively. Note that if at the time of programming a WL, a cell is kept erased,  then $Y_{j}^{a}=Y_j^{a,e}$ and hence the erased cell does not interfere with other cells. Due to the variability of the manufacturing process, a pair of cells may have a different coupling than that of another pair. However, the effect of this variability is negligible compared to the voltage shift $\left(Y_{j}^{a}-Y_j^{a,e}\right)$ with which it is multiplied. This shift is at least one order of magnitude larger. Therefore, to maintain mathematical tractability, this variability is ignored, and cells separated by the same distance are assumed to have the same coupling ratio. In particular, the cells that are direct neighbors (same WL/BL or adjacent WL/BL) in the vertical, diagonal directions, have coupling ratios $\gamma_v$ and $\gamma_d$ respectively, independently of their position in the block.

\section{The $\text{V}_{th}$ Distribution of a Single Cell}
\label{SingleCellDist}
In this section, we derive the distribution of ICI in a victim cell and the distribution of the output of a single cell given its input. The input levels to all cells are assumed independently and uniformly distributed (i.u.d).

The ICI expression of  \eqref{eq:ICIexp} can be expanded using \eqref{channelmodel} as follows 
\begin{align}
\label{eq:ICIexpanded}
W=&\sum_{j=1}^{m}\gamma_{j}\left(v_{X_j^a}+Z_j^{a}-v_0-Z_j^{a,e}\right),
\end{align}
where $Z_j^{a,e}$ and $Z_j^{a}$ are the noise that the $j$th  aggressor suffer from before and after being programmed, and $X_j^a$ is the input to the $j$th aggressor. Therefore, ICI is a linear combination of the shifts in the aggressor cells due to their programming. Note that \eqref{eq:ICIexpanded} assumes that since WLs are programmed in order, the aggressor cell does not originally suffer from ICI when it subjects the victim cell to ICI, and that any subsequent ICI that the aggressor is subject to, does not affect the original victim.
It is easy to see that given $X_j^a$, $\left(v_{X_{j}^a}+Z_j^a-v_{0}-Z_j^{a,e}\right)$ follows a Gaussian distribution (difference of two independent Gaussians). It is easy to see that given the inputs to the aggressors, which we denote by vector $\mathbf{X}^a=(X_1^a,....,X_m^a)$, $W$ is a linear combination of independent Gaussian variables which is also Gaussian. Therefore, we can write that
\begin{equation}
p_{W|\mathbf{X}^a}(w|\mathbf{x}^a)=\mathcal{N}\Big(w|\psi(\mathbf{x}^a),\theta^2(\mathbf{x}^a)\Big),
\end{equation}
where 
\begin{align}
\psi(\mathbf{x}^a)=&\sum_{j=1}^m\gamma_j(v_{x_j^a}-v_0),\\
\theta^2(\mathbf{x}^a)=&\sum_{j=1}^m\gamma_j^2\big(1-\delta_{x_j^a,0}\big)(\sigma_{x_j^a}^2+\sigma_{0}^2),
\end{align}
where the Kronecker-delta indicator reflects the fact that when a cell is kept erased, no ICI-free shift occurs in it.
Given $X$, $Z$ also follows a Gaussian distribution. Thus, given the inputs of the victim and aggressors, $Z+W$ is a sum of two Gaussian variables, and
\begin{equation}
p_{Y|X,\mathbf{X}^a}(y|x,\mathbf{x}^a)=\mathcal{N}\Big(y|v_x+\psi(\mathbf{x}^a),\sigma_x^2+\theta^2(\mathbf{x}^a)\Big).
\end{equation}
 The conditional distribution of the output given its input is then obtained by marginalization as
 \begin{align}
\label{Eq:SingleCellDist}
 p_{Y|X}(y|x)=&\sum_{\mathbf{x}^a} p_{\mathbf{X}^a|X}(\mathbf{x}^a|x)p_{Y|X,\mathbf{X}^a}(y|x,\mathbf{x}^a)\nonumber\\
 =&\frac{1}{q^m}\sum_{\mathbf{x}^a}\mathcal{N}\Big(y|v_x+\psi(\mathbf{x}^a),\sigma_x^2+\theta^2(\mathbf{x}^a)\Big),
 \end{align}
 where the second equality is due to the input of all cells being i.u.d. For the case of ABL discussed in section \ref{flashBasics}, $m=3$ and the Gaussian mixture contains $q^3$ components.

Equation \eqref{Eq:SingleCellDist} shows that the read $\text{V}_{th}$ of a cell given its input  follows a Gaussian mixture distribution with one component per realization of the inputs of the aggressors. This result coincides with the experimental results of \cite{cai2013program} where a similar behaviour was observed.

\section{Model-based Detector}
\label{GraphMod}
\subsection{Graphical Model}
In this section, instead of considering a single victim cell, we consider a group of $n$ victim cells that constitute a WL.  We denote by $\mathbf{X}=\left(X_1,...,X_n\right)^T\in\mathbb{Z}_q^n$ and  $\mathbf{Y}=\left(Y_1,...,Y_n\right)^T\in \mathbb{R}^n$ the inputs and outputs of the victim cells, and we are interested in deriving an accurate approximation of the conditional joint distribution of the outputs of the victim cells given their inputs, namely $p_{\mathbf{Y}|\mathbf{X}}(\mathbf{y}|\mathbf{x})$. Note that an exact expression of $p_{\mathbf{Y}|\mathbf{X}}(\mathbf{y}|\mathbf{x})$ can be obtained through a similar approach to that of the previous section, but it results in a multi-variate Gaussian mixture with $q^n$ components. Given that $n$ is typically on the order of thousands, this exact model is prohibitively complex to implement in practice for detection purposes.

\begin{figure}[t]
\centering
\includegraphics[ scale=0.8]{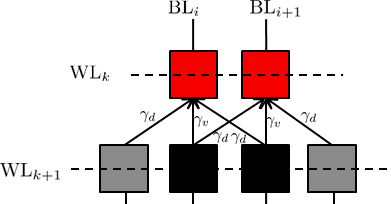}
\caption{Victim word-line (red) and agressor word-line (grey).}
\label{AnaStruct}
\end{figure}

Since noise is independent from one cell to another, it does not introduce any correlation between them. However, even when  two cells are not interfering with each other (as is the case for cells on the same WL in ABL), the fact that an aggressor can subject more than a single cell to ICI will induce a correlation between them.  Fig. \ref{AnaStruct} shows (in black) the common aggressors, due to diagonal coupling, of two adjacent cells (in red). Consequently, two adjacent victim cells indexed by $i$ and $i+1$ will be subject to a correlated ICI, i.e., $W_{i}$ and $W_{i+1}$ are not independent. 

The ICI expression in \eqref{eq:ICIexpanded} can be re-written as 
\begin{align}
\label{Eq:ICIcontdisc}
W= &\sum_{j=1}^{m}\gamma_{j}\left(v_{X_j^a}-v_0\right)+\sum_{j=1}^{m}\gamma_{j}\left(Z_j^{a}-Z_j^{a,e}\right).
\end{align}
where the leftmost sum is a discrete interference term caused by the ideal shift in the aggressor cell, and the rightmost sum is a continuous propagated noise term that the interfering cells are subject to. Subsequently,  the channel model of \eqref{channelmodel} can be re-written as follows
\begin{equation}
\label{channelmodelNew}
Y= v_X+\sum_{j=1}^{m}\gamma_{j}\left(v_{X_j^a}-v_0\right)+\underbrace{Z+\sum_{j=1}^{m}\gamma_{j}\left(Z_j^{a}-Z_j^{a,e}\right)}_{Z'}.
\end{equation}
where $Z'$ is an equivalent noise term that includes the direct and propagated noise in the victim cell. However, in contrast to $Z$ which was independent from one cell to another, $Z'$ is correlated since it depends on the noise in the aggressors, which as discussed above, may be shared among victims. 

\begin{figure}[t]
\centering
\includegraphics[ scale=0.7]{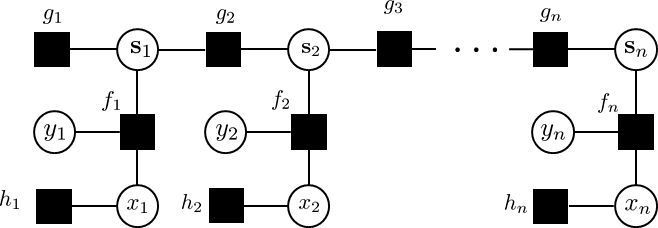}
\caption{Factor Graph representing \eqref{eq:finalFactorization}}
\label{FactGraph}
\end{figure}

In order to obtain a mathematically tractable form, we assume that this new term is independent from one cell to another. As opposed to when the whole of ICI is assumed independent from one cell to another, the outputs of adjacent cells will still be correlated in our approximation due to the discrete term in  \eqref{channelmodelNew}. To summarize our approximation: in the actual model, the output of a cell depends on its input and the \emph{shifts} in its aggressor, which are continuous. In our approximated model, the output of a cell depends on its input and the \emph{inputs} of its aggressors, which are discrete. If we define the state of a cell by the inputs of its aggressor cells and we denote it by $\mathbf{S}_i=(X_{i-1}^a,X_{i}^a,X_{i+1}^a)$, the channel reduces to the well known finite state Markov channel (FSMC) \cite{goldsmith1996capacity} with 

\begin{equation}
p_{\mathbf{Y}|\mathbf{X},\boldsymbol{S}}(\mathbf{y}|\mathbf{x},\boldsymbol{s})=\prod_{i=1}^np_{Y|X,\mathbf{S}}(y_i|x_i,\mathbf{s}_i)
\end{equation}
 where $\boldsymbol{S}=(\mathbf{S}_1,...,\mathbf{S}_n)$, and
 \begin{equation}
 p_{Y|X,\mathbf{S}}(y_i|x_i,\mathbf{s}_i)=\mathcal{N}\Big(y_i|v_{x_i}+\psi(\mathbf{s}_i),\sigma_i^2+\theta^2(\mathbf{s}_i)\Big).
\end{equation}
From a graphical modeling point of view, the joint distribution $p_{\mathbf{Y},\mathbf{X},\boldsymbol{S}}(\mathbf{y},\mathbf{x},\boldsymbol{s})$ can be factorized as follows
\begin{multline}
\label{eq:finalFactorization}
p_{\mathbf{X},\mathbf{Y},\boldsymbol{S}}(\mathbf{x},\mathbf{y},\boldsymbol{s})=\prod_{i=1}^{n}\underbrace{p_{X}(x_{i})}_{h_{i}\left(x_{i}\right)}\underbrace{p_{\mathbf{S}_{i}|\mathbf{S}_{i-1}}(\mathbf{s}_{i}|\mathbf{s}_{i-1})}_{g_{i}\left(\mathbf{s}_{i},\mathbf{s}_{i-1}\right)}\\
\underbrace{p_{Y|X,\mathbf{S}}(y_{i}|x_{i},\mathbf{s}_{i})}_{f_{i}\left(y_{i},x_{i},\mathbf{s}_{i}\right)},
\end{multline}
where the sequences of functions $f_i,g_i$ and $h_i$ are introduced for clarity of representation, and the variables of each function will be omitted but should be clear from its index. Since adjacent cells have common aggressors, their states are not independent and form a Markov chain where $p_{\mathbf{S}_i|\mathbf{S}_{i-1}}(\mathbf{s}_i|\mathbf{s}_{i-1})$ represents the dependency between the states of adjacent cells. In particular, the two leftmost components of $\mathbf{S}_i$ are always identical to the two rightmost  components of $\mathbf{S}_{i-1}$, while the rightmost component of  $\mathbf{S}_i$ follows the input distribution, which is i.u.d. In other words, $p_{\mathbf{S}_i|\mathbf{S}_{i-1}}(\mathbf{s}_i|\mathbf{s}_{i-1})=0$ if $(s_{i,1},s_{i,2})\neq(s_{i-1,2},s_{i-1,3})$ and $\frac{1}{q}$ otherwise.  The  factorization in \eqref{eq:finalFactorization} can be represented by the factor graph \cite{kschischang2001factor} shown in Fig. \ref{FactGraph}, where circles represent variables and squares represent functions. 

It is worth noting that according to the above model, the flash memory channel in the presence of ICI is equivalent to a FSMC \cite{goldsmith1996capacity} where the state of the channel changes according to a finite-state Markov process. In this analogy, a cell corresponds to a discrete time-step. The FSMC is defined by its conditional input/output probability at cell $i$ by $p_{Y|X,\mathbf{S}}(y_{i}|x_{i},\mathbf{s}_{i})$. This distribution in our case is Gaussian with mean and variance given by
\begin{align}
\label{Mean}
\mu_{x,\mathbf{s}}&=\mathbb{E}\left[Y|X=x,\mathbf{S}=(s_1,s_2,s_3)\right]\nonumber \\
&=v_{x}+\sum_{i=1}^{3}\gamma_i\left(v_{s_{i}}-v_{0}\right),\text{and}
\end{align}
\begin{align}
\label{Variance}
\sigma_{x,\mathbf{s}}^{2}&=\mathbb{V}\left[Y|X=x,\mathbf{S}=(s_1,s_2,s_3)\right] \nonumber\\
&=\sigma_{x}^{2}+\sum_{i=1}^{3}(1-\delta_{s_{i},0})\gamma_{i}^{2}\left(\sigma_{s_{i}}^{2}+\sigma_{0}^{2}\right),
\end{align}
where $\delta_{i,j}$ is the Kronecker delta function (erased cells do not interfere), $\gamma_2=\gamma_v$ and $\gamma_1=\gamma_3=\gamma_{d}$. Therefore, the  channel is Gaussian but with variable mean and variance that depend on the state of the cell which itself depends on the states of adjacent cells.

\subsection{Detector}
\begin{algorithm}
\caption{Posterior Calculation Using Message-Passing}\label{SPAlgo}
\begin{algorithmic}[1]
\For{$i=1:1:n$}
\State $\mu_{h_i\rightarrow X_i}(x_i)=p(x_i)$
\State $\mu_{X_i\rightarrow f_i}(x_i)=\mu_{h_i\rightarrow X_i}(x_i)$
\State $\mu_{f_i\rightarrow \mathbf{S}_i}(\mathbf{s}_i)=\sum_{X_i}p\left(y_i|x_i,\mathbf{s}_i\right)\mu_{X_i\rightarrow f_i}(x_i)$
\EndFor\label{euclidendfor}

\State $\mu_{g_1\rightarrow \mathbf{S_1}}(\mathbf{s}_1)=p(\mathbf{s}_1)$
\For{$i=1:1:n-1$} \Comment{Forward Messages}
\State $\mu_{\mathbf{S}_i\rightarrow g_{i+1}}(\mathbf{s}_i)=\mu_{f_i\rightarrow \mathbf{S}_i}(\mathbf{s}_i)\mu_{g_i\rightarrow \mathbf{S_i}}(\mathbf{s}_i)$
\State $\mu_{g_{i+1}\rightarrow \mathbf{S}_{i+1}}(\mathbf{s}_{i+1})=\sum_{\mathbf{s}_i}p\left(\mathbf{s}_{i+1}
|\mathbf{s}_i\right)\mu_{\mathbf{s}_i\rightarrow g_{i+1}}(\mathbf{s}_{i})$
\EndFor\label{ForwardMessages}

\State $\mu_{\mathbf{S}_n\rightarrow g_n}(\mathbf{s}_{n})=\mu_{f_n\rightarrow \mathbf{S}_n}(\mathbf{s}_{n})$

\For{$i=n-1:1:1$} \Comment{Backward Messages}
\State $\mu_{g_{i+1}\rightarrow \mathbf{S}_{i}}(\mathbf{s}_{i})=\sum_{\mathbf{s}_{i+1}}p\left(\mathbf{s}_{i+1}
|\mathbf{s}_{i}\right)\mu_{\mathbf{S}_{i+1}\rightarrow g_{i+1}}(\mathbf{s}_{i+1})$
\State $\mu_{\mathbf{S}_{i}\rightarrow g_{i}}(\mathbf{s}_{i})=\mu_{f_{i}\rightarrow \mathbf{S}_{i}}(\mathbf{s}_{i})\mu_{g_{i+1}\rightarrow \mathbf{S_i}}(\mathbf{s}_{i})$
\EndFor\label{BackwardMessages}

\For{$i=1:1:n$}
\State $\mu_{ \mathbf{S}_i \rightarrow f_i}(\mathbf{s}_{i})=\mu_{g_i\rightarrow \mathbf{S}_i}(\mathbf{s}_{i})\mu_{g_{i+1}\rightarrow \mathbf{S}_i}(\mathbf{s}_{i})$
\State $\mu_{f_i\rightarrow X_i}(x_{i})=\sum_{\mathbf{s}_i}p(y_i|x_i,\mathbf{s}_i)\mu_{ \mathbf{S}_i \rightarrow f_i}(\mathbf{s}_{i})$
\State $p(x_i|\mathbf{y})\propto\mu_{h_i\rightarrow X_i}(x_i)\mu_{f_i\rightarrow X_i}(x_i)$
\EndFor\label{Marginals}

\end{algorithmic}
\end{algorithm}

When the access to the outputs or inputs of the aggressor WL is prohibited, cancellation or equalization is not possible. Therefore, the detector can exploit only the outputs $\mathbf{Y}$ of the victim cells to estimate the inputs. The soft detector is obtained by calculating the posterior probability mass function (PMF) of each cell, denoted by $p_{X_i|\mathbf{Y}}(x|\mathbf{y})$ for $i=1,...,n$, while the optimal hard detector is obtained by maximizing these PMFs to minimize the symbol error rate, i.e., 
\begin{equation}
\label{Eq:ReconstructionFunc}
x_i^{*}=\argmax_{x\in \mathbb{Z}_q} p_{X_i|\mathbf{Y}}(x|\mathbf{y}).
\end{equation}
where $x_i^{*}$ is the optimal estimate of the symbol stored in the $i$th cell. Both hard and soft detection can be passed on to hard or soft decoder respectively.
The PMF $p_{X_i|\mathbf{Y}}(x|\mathbf{y})$ is proportional to  $p_{X_i,\mathbf{Y}}(x,\mathbf{y})$, which is the marginal of $p_{\mathbf{X},\mathbf{Y},\underbar{S}}(\mathbf{x},\mathbf{y},\underbar{s})$. The marginalization is done over all unknown variables except $x_i$, i.e., 
\begin{align}
\label{Eq:MarginGen}
p_{X_i,\mathbf{Y}}(x,\mathbf{y})= \sum_{\sim \{x_i\}}p_{\mathbf{X},\mathbf{Y},\underbar{S}}(\mathbf{x},\mathbf{y},\underbar{s})
\end{align}
where  $\sum_{\sim \{x_i\}}$  is the summation over all unknown variables except $x_i$. This marginalization problem can be solved by applying the sum-product algorithm (shown in Algorithm \ref{SPAlgo}) to the factor graph representation of Fig. \ref{FactGraph}. 
Since the corresponding factor graph is a tree, the sum-product algorithm provides the exact marginals for all variables by running the forward and backward recursions once without the need for any iterations. The messages $\mu_{a\rightarrow b}$ and $\mu_{b\rightarrow a}$ are exchanged between each pair of adjacent nodes $a$ and $b$  according to Algorithm \ref{SPAlgo} where the messages are always functions of the variable involved \cite{kschischang2001factor}. In practice, appropriate normalization of the messages is required to ensure their values do not rapidly go to zero for large $n$. Note that the complexity of the algorithm is linear in $n$.

\begin{figure*}[t]
    \begin{subfigure}[t]{0.51\textwidth}
    \centering
\includegraphics[ scale=0.5]{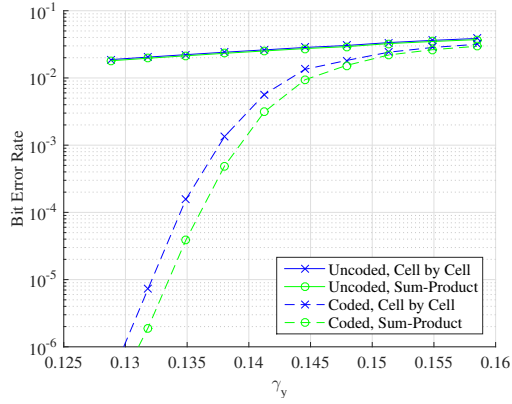}
\caption{$\beta=1,\alpha=0.075$.}
\label{Fig:HMMPerf1}
    \end{subfigure}
    \begin{subfigure}[t]{0.5\textwidth}
\centering
\includegraphics[ scale=0.5]{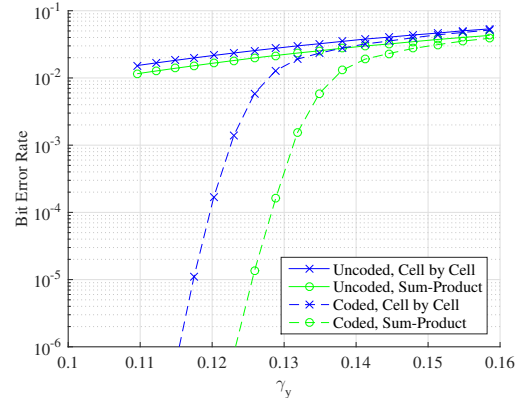}
\caption{$\beta=1,\alpha=0.25$.}
\label{Fig:HMMPerf2}
    \end{subfigure}
    \begin{subfigure}[t]{0.51\textwidth}
\centering
\includegraphics[ scale=0.5]{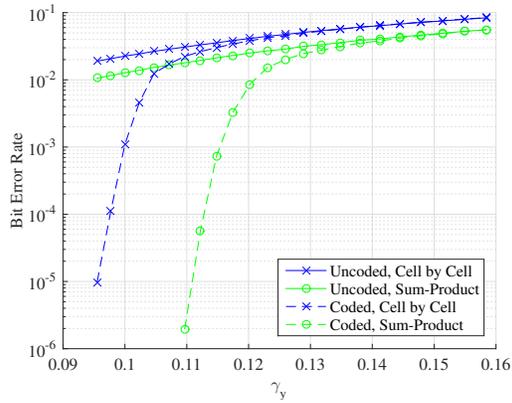}
\caption{$\beta=1,\alpha=0.5$.}
\label{Fig:HMMPerf3}
    \end{subfigure}
    \begin{subfigure}[t]{0.5\textwidth}
\centering
\includegraphics[ scale=0.5]{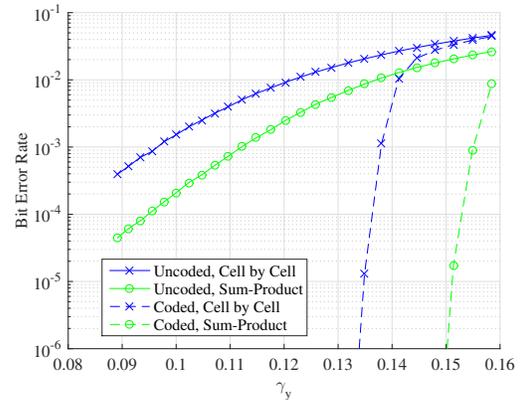}
\caption{$\beta=0.5,\alpha=0.25$.}
\label{Fig:HMMPerf4}
    \end{subfigure}
    \caption{Coded and uncoded BER using of the cell by cell detector (blue) and the model-based detector (in green) as a function of $\gamma_v$ and for different $\beta$ and $\alpha$.}
\label{Fig:CompareJSHMM}
\end{figure*}

\section{Simulation Results}
\label{SImRes}
 To demonstrate the benefits of the proposed detector, we compare its coded and uncoded performances against the benchmark.
As a benchmark detector, we consider the sub-optimal cell by cell detector where each cell  is detected based only on its output, i.e., the soft detector is simply obtained by evaluating  $p_{X|Y}(x|y_i)$ for all $i$ and the hard detector is obtained by maximizing these posteriors,
where $p_{X|Y}(y|x)\propto p_{Y|X}(y|x)$ which was derived in \eqref{Eq:SingleCellDist} of section \ref{SingleCellDist}. This sub-optimal detector, which is often employed in the literature, becomes optimal only when the cells are actually independent, i.e., diagonal interference is negligible.
 
 For the uncoded performance, the posteriors are simply maximized while, for the coded one, the posterior PMF (or the log-likelihood ratios evaluated from it) of each cell is passed to the soft-LDPC decoder. We simulate an MLC device ($q=4$) and employ the simulation parameters given in Table \ref{Parameters1}, where $\beta$ controls the strength of noise in the system. The channel parameters are assumed to be perfectly known and are given for each input/state combination by \eqref{Mean} and \eqref{Variance}.
 
We plot the bit error rate performance as a function of the vertical coupling $\gamma_v$ and for different values of the diagonal coupling $\gamma_d$, which we set as a fraction $\alpha$ of the vertical coupling, i.e., $\gamma_d=\alpha \gamma_v$. The outputs of the cells are assumed to be available with floating point precision, i.e., unquantized. We use a rate-$0.89$, length-$9216$ LDPC code and the built-in Matlab encoder and decoder with the maximum number of iterations set to $50$. Simulations were run until the decoded WL was erroneous in at least 100 Monte Carlo trials?
\begin{table}
\center
\begin{tabular}{|c|c|c|c|c|}
\hline 
$x$ & $0$ & $1$ & $2$ & $3$\tabularnewline
\hline 
\hline 
$v_{x}$ & $1.1$ & $2.7$ & $3.3$ & $3.9$\tabularnewline
\hline 
$\sigma_{x}$ & $0.35\beta$ & $0.09\beta$ & $0.09\beta$ & $0.09\beta$\tabularnewline
\hline 
\end{tabular}
\caption{Input and noise parameters used for simulation.}
\label{Parameters1}
\end{table}

The results are shown in Fig. \ref{Fig:CompareJSHMM}. As mentioned before, when there is no diagonal coupling, i.e., $\alpha=0$, the two detectors are equivalent and hence provide the same performance (the plots were omitted). As the diagonal coupling starts to increase, i.e., $\alpha=0.075$, the uncoded performance of the two detectors is virtually the same, while the proposed detector provides an improvement in the coded performance over the cell by cell detector; e.g., at $\gamma_v=0.135$, the proposed scheme reduces the BER by $75\%$. The benefit of sum-product detector becomes more significant for $\alpha=0.25$, especially in the coded case where we can see at $\gamma_v=0.126$ a $99.7\%$ reduction. 
For extreme cases of diagonal coupling ($\alpha=0.5$), the uncoded BER experience up to $44.5\%$ reduction at $\gamma_v=0.0955$, while the minimum reduction in the coded BER is approximately $34\%$ $(\gamma_v=0.1585)$ and reduction can reach  up to $99.99\%$ ($\gamma_v<0.1096$). 
In general, the results show that the proposed detector provides more accurate soft-input to the LDPC decoder by exploiting the dependence between adjacent cells due to common aggressors. 

%
 
The choice of $\beta=1$ reflects a significant level of noise in the system. To see what happens when the system is less affected by noise, we show in Fig. \ref{Fig:HMMPerf4} the same results for $\beta=0.5$ and $\alpha=0.25$. The gap between the performance of the two detector widens for both coded and uncoded performance. This shows that in cases where the system is limited by interference, exploiting the dependency between adjacent cells can bring significant improvements to the performance. 
As devices scale down, they become more and more limited by ICI. By having a more accurate posterior PMF, the coded and uncoded performance can be significantly improved.

 \section{Effect of Quantization on the Proposed Detector}
 \label{EffectQuantization}
In the previous section, it was assumed that the output of a cell is available in floating point precision, i.e., unquantized. While high-precision is possible in practice, there is a trade-off between reading latency and output precision as explained in section \ref{flashBasics}. Therefore, it is desirable to have a small number of required sensings to achieve an acceptable reliability level. In this section, we assess the effect of having a low-precision output on the inference performance. We show that when the output is quantized, the performance of the proposed sum-product detector deteriorates significantly. Therefore, we propose an iterative scheme to partially recover this drop in performance.

\begin{figure*}[t]
    \begin{subfigure}[t]{0.51\textwidth}
    \centering
\includegraphics[ scale=0.5]{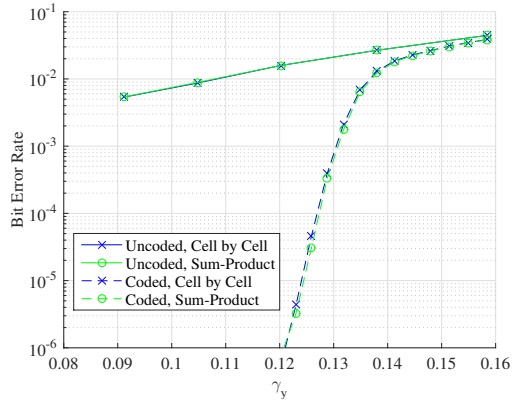}
\caption{$\beta=1,\alpha=0.075$.}
\label{Fig:HMMPerf1Q}
    \end{subfigure}
    \begin{subfigure}[t]{0.5\textwidth}
\centering
\includegraphics[ scale=0.5]{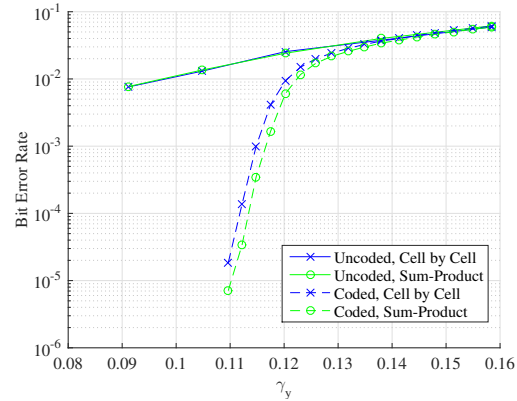}
\caption{$\beta=1,\alpha=0.25$.}
\label{Fig:HMMPerf2Q}
    \end{subfigure}
    \begin{subfigure}[t]{0.51\textwidth}
\centering
\includegraphics[ scale=0.5]{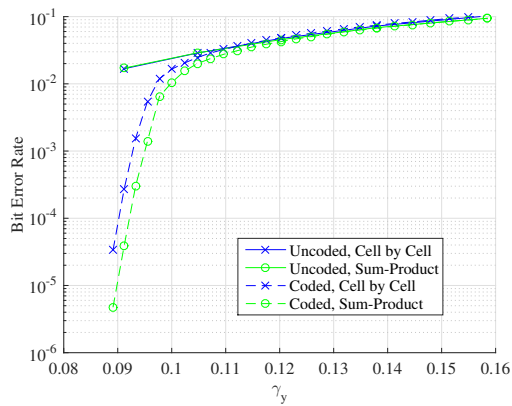}
\caption{$\beta=1,\alpha=0.5$.}
\label{Fig:HMMPerf3Q}
    \end{subfigure}
    \begin{subfigure}[t]{0.5\textwidth}
\centering
\includegraphics[ scale=0.5]{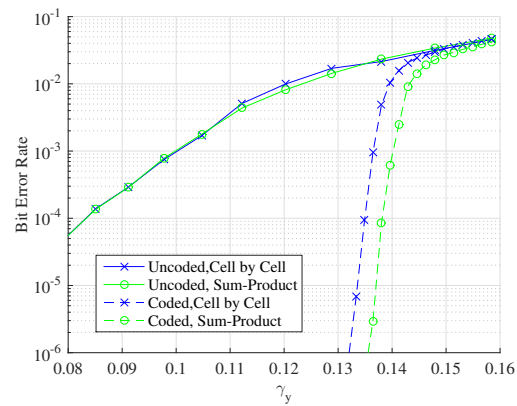}
\caption{$\beta=0.5,\alpha=0.25$.}
\label{Fig:HMMPerf4Q}
    \end{subfigure}
    \caption{Coded and uncoded BER using of the cell by cell detector (blue) and the model-based detector (in green) when the output is quantized as a function of $\gamma_v$ and for different $\beta$ and $\alpha$.}
\label{Fig:CompareJSHMMQ}
\end{figure*}
\subsection{Quantization}
\begin{figure}[t]
    \centering
\includegraphics[ scale=0.5]{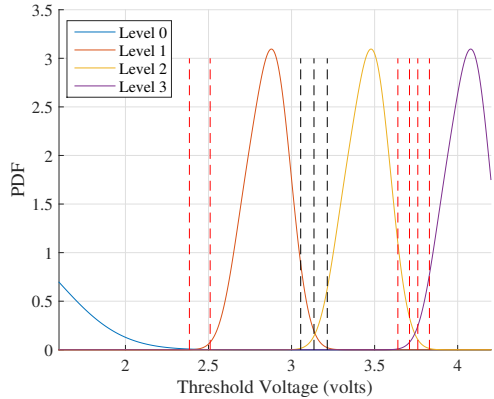}
\caption{The quantization boundaries are shown in black when $\beta=1$, $\gamma_v=0.08$ and $\alpha=0.075$.}
\label{Fig:Quantizer}
    \end{figure}

Several quantization schemes were proposed to improve soft decoding of LDPC codes. In \cite{dong2011use}, it was shown that non-uniform quantization improves the performance of soft decoded LDPC codes relative to uniform quantization. In \cite{wang2011soft}, it was shown that a quantizer that maximizes the mutual information between the input and the quantized output improves soft decoding. At the intuitive level, the quantized output should preserve as much information about the input as possible. Since in practice, the LSB and MSB pages are read independently, we consider three and six read references to read the LSB  and MSB pages, respectively. Fig. \ref{Fig:Quantizer} shown an example of the quantization boundaries used for detecting the LSB (in black) and the MSB (in red) when $\beta=1$, $\gamma_v=0.08$ and $\alpha=0.075$. The references for each page were obtained such as to maximize the mutual information between the binary input of the page and the quantized output, i.e.,
\begin{align}
\label{Ed:SubOptQuantizer}
\textit{Q}_L^*&=\argmax_{\textit{Q}\in \mathcal{Z}_4} I\big(X_L;\textit{Q}(Y)\big)\\
\textit{Q}_M^*&=\argmax_{\textit{Q}\in \mathcal{Z}_7} I\big(X_M;\textit{Q}(Y)\big),
\end{align}
where $X_L\in\{0,1\}$ and $X_M\in\{0,1\}$ are the LSB and MSB bits of a cell, respectively,  $\textit{Q}_L^*$ and $\textit{Q}_M^*$ are the LSB and MSB quantizers, respectively. $\mathcal{Z}_{k}$ is the set of all scalar, deterministic quantizers with $k$ outputs, and $\textit{Q}:\mathbb{R}\rightarrow \mathbb{Z}_{k}$ is a many-to-one function in $\mathcal{Z}_{k}$ that maps the real output to a quantized output in  $\mathbb{Z}_{k}$. 

In \cite{kurkoski2014quantization}, a low complexity algorithm to obtain such quantizers for binary-input discrete memoryless channels was proposed, which we adopt. However, it is important to note that this algorithm assumes that the channel is memoryless, therefore, the obtained quantizers are optimal only when the input of each cell is inferred using its output, which is true only for our benchmark. However, in the case of the proposed detector, the input of a cell is inferred from the outputs of a group of cells, and hence, it is not guaranteed that such quantizer maximizes the mutual information between the input and the quantized outputs, i.e., $I\big(X_L;Q(Y_1),...,Q(Y_N)\big)$ and $I\big(X_M;Q(Y_1),...,Q(Y_n)\big)$. Nevertheless, we do not aim to design an optimal quantizer, and the performance of the sum-product approach for this quantizer can be considered as a lower bound on the optimal performance. 

\subsection{Performance Under Quantization}
\label{sec:MaxOut1}

The performances of both detectors under quantization are shown in Fig. \ref{Fig:CompareJSHMMQ}. Despite still being superior, the performance of the sum-product detector is more negatively affected by quantization compared to cell by cell detector. In the uncoded case, the performance does not exhibit any improvement, even in the ICI-limited case. The coded performance on the other hand exhibits an improvement that is small compared to the improvement observed in the unquantized case. As an example, for $\beta=1,\alpha=0.25$ and $\gamma_v=0.126$, the reduction in the coded BER is only $14.1\%$ in the quantized case compared to $99.7\%$ in the unquantized case. This deterioration is accompanied with a general shift to the left of the plots, indicating that quantization significantly increase the BER for both detector, which is expected.

This significant sensitivity of the proposed detector can be explained by the fact that quantization, especially when it has low-precision, discards some information from the output signal. The sensitivity to quantization is then a matter of how much the inference scheme has been initially benefiting from the discarded information. In our case, the sum-product benefited significantly more than the benchmark from the outputs when they were not quantized to learn  information about ICI and provide high-quality inputs to the soft decoder. The quantized output contains information about the input, noise, and ICI. When the noise is significant, a big part of the information kept due to quantization will be related to noise, which is independent, and the performance of the sum-product seems to become very similar to that of the benchmark. However, when the system becomes limited by ICI, e.g., $\beta=0.5$ as shown in Fig. \ref{Fig:HMMPerf4Q}, it is clear that the quantized output carries much more information about ICI, which can now be exploited more efficiently by the sum-product, and we can see the performance gap between the sum-product and the benchmark partially recovering.

As a conclusion, the proposed sum-product algorithm is sensitive to quantization and can cope better with it in an ICI-limited scenario. As explained before, the quantization scheme that we employ is not necessarily optimal for our detector, and it may be possible to design a better quantizer that extracts more information to inform the sum-product algorithm. However, in the next section, we use a different approach to improve the performance of the sum-product algorithm using the same quantization scheme.

\begin{figure*}[t]
    \begin{subfigure}[t]{0.51\textwidth}
    \centering
\includegraphics[ scale=0.5]{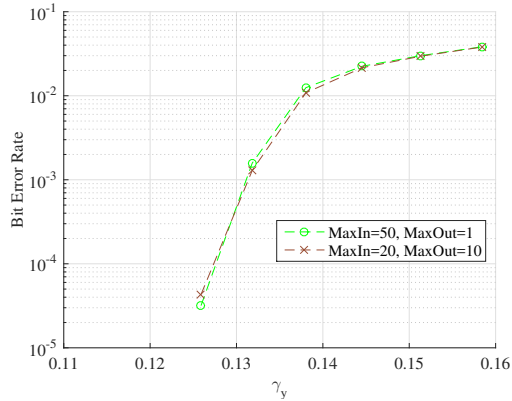}
\caption{$\beta=1,\alpha=0.075$.}
\label{Fig:AlterQ1}
    \end{subfigure}
    \begin{subfigure}[t]{0.5\textwidth}
\centering
\includegraphics[ scale=0.5]{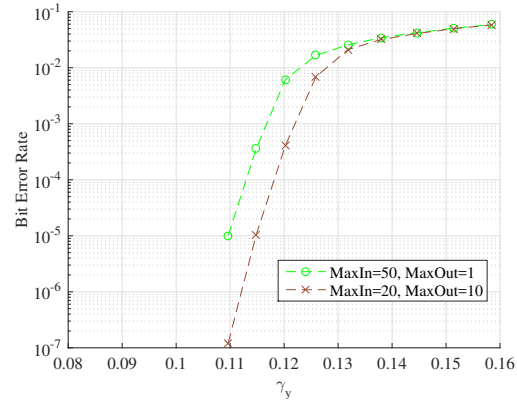}
\caption{$\beta=1,\alpha=0.25$.}
\label{Fig:AlterQ2}
    \end{subfigure}
    \begin{subfigure}[t]{0.51\textwidth}
\centering
\includegraphics[ scale=0.5]{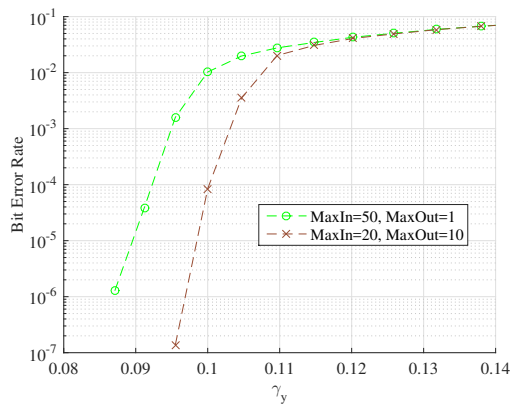}
\caption{$\beta=1,\alpha=0.5$.}
\label{Fig:AlterQ3}
    \end{subfigure}
    \begin{subfigure}[t]{0.5\textwidth}
\centering
\includegraphics[ scale=0.5]{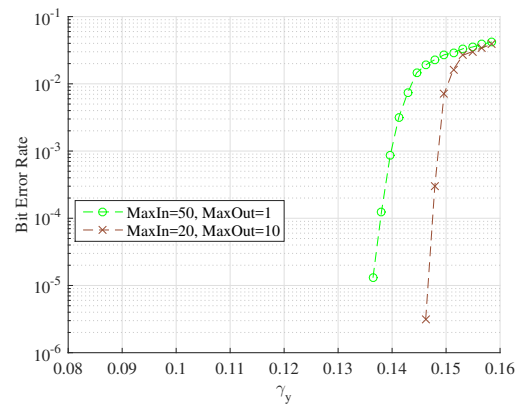}
\caption{$\beta=0.5,\alpha=0.25$.}
\label{Fig:AlterQ4}
    \end{subfigure}
    \caption{Coded BER using the sum-product with (green) and without (brown) the iterative scheme. $MaxIn$ is the maximum number of iterations that the decoder performs and $MaxOut$ is the maximum number of times the detection/decoding process is done.}
\label{Fig:AlterQ}
\end{figure*}

\subsection{Iterative Detection/Decoding Strategy}
In our setting, the sum-product algorithm is used to soft detect the input symbols and subsequently pass the soft information to the soft decoder. In other words, the detector informs the decoder but not vice-versa. The structure of the ECC obtained from the constraint $H\mathbf{x}= \boldsymbol{0}^T$ can be represented by  a factor graph (Tanner graph)\cite{kschischang2001factor}. Therefore, the channel and code factor graphs can be combined into a single factor graph representing the system on which inference is to be performed using the sum-product algorithm. In this case, no differentiation is made between detection and decoding. 

As a middle ground, we propose a strategy that still differentiates between detection and decoding, but allows the detector to inform the decoder and vice-versa, by iterating multiple times between detection and decoding. In other words, the output of the decoder, which is another posterior distribution on the input, is passed back to the detector as a new prior, after removing from it the intrinsic information. This is equivalent to iterative receivers used for FSMC as in \cite{ratzer2002low}. Note that the decoder itself is an iterative instance of the sum-product since it has cycles, while the detector is a cycle-free instance of it. Hence, we differentiate between the number of inner iterations of the decoder, and the number of outer iterations that switches between decoding and detecting. We denote the maximum number of these iterations as $MaxIn$ and $MaxOut$, respectively. The scheme evaluated in the previous sections corresponds to $MaxIn=50, MaxOut=1$ which means there is no iteration between detection and decoding. Fig. \ref{Fig:AlterQ} shows the results when the output is quantized and when the iterative strategy between detection and decoding is employed with $MaxIn=20$ and $MaxOut=10$ and compares it to the case of the previous subsection. The choice of the of $MaxIn=20$ $MaxOut=10$ was made arbitrarily based on a few trials. From the figure, it is clear that the proposed strategy can mitigate the effects of quantization and the performance is improved significantly especially for larger diagonal coupling. Note that even in the unquantized case of Fig. \ref{Fig:CompareJSHMM}, the significant improvement was for larger diagonal coupling as well. When the system is ICI-limited as shown in \ref{Fig:AlterQ4}, the iterative strategy significantly improves the performance by several orders of magnitude.

\section{Conclusion and Future Work}
\label{Conclusion}
This paper presented a model-based detector that enables soft LDPC decoder to perform significantly better in the presence of ICI. The work is particularly relevant for the highly parallel SSDs of the future which will be based on significantly scaled flash chips. In these devices, powerful, yet simple signal processing techniques will be a key factor contributing to their mass adoption. We have shown that the proposed detector is sensitive to low-precision output. To overcome this, we applied the concept of joint detection and decoding and proposed an iterative scheme that alternates between detection and decoding multiple times to recover the data.  This scheme was shown to significantly reduce the adverse effect of quantization.

The work demonstrated the potential of joint detection of cells, as opposed to the state-of the art schemes that either detect cells individually or employ expensive ICI cancellation. This opens the door for several future works such as the extension to different architectures and different channel models. As the work assumed perfect knowledge of channel parameters, future work may study efficient methods to estimate these parameters given the proposed model. Furthermore, since we identified the finite-state Markov channel model as a more accurate model of flash memory, it will open the door to answer theoretical and practical questions about the information theoretic capacity of flash memory\cite{goldsmith1996capacity} and the design of suitable LDPC codes \cite{eckford2007designing}.





%
 \bibliographystyle{ieeetr}
 \bibliography{ref}

\end{document}